\documentclass[11pt]{article}

\usepackage{graphicx,epsf,subfigure,rotate,epsfig}
\usepackage{pstricks,pst-node,psfrag}
\usepackage{amsthm,amssymb,amsmath}
\usepackage{bm}
\usepackage{hyperref}
\usepackage{booktabs}
\usepackage{setspace}
\usepackage{latexsym}
\usepackage{lineno,natbib}
\usepackage{rotating}

\def\keywords{\vspace{.5em}
{\textit{Keywords}:\,\relax%
}}

\newcommand{\eps}{\epsilon}
\newcommand{\beps}{\bm {\epsilon}}
\newcommand{\bmeta}{\bm \eta}
\newcommand{\half}{\frac12}
\newcommand{\diag}{\text{diag}}
\newcommand{\mN}{\mbox{N}}
\newcommand{\bSigma}{\bm \Sigma}
\newcommand{\bP}{\bm P}

\newcommand{\bS}{\bm S}
\newcommand{\bA}{\bm A}
\newcommand{\bB}{\bm B}

\newcommand{\bOmega}{\bm \Omega}
\newcommand{\bmu}{\bm \mu}
\newcommand{\bX}{\bm X}
\newcommand{\etr}{\text{etr}}
\newcommand{\tr}{\text{trace}}

\newcommand{\bY}{\bm Y}
\newcommand{\be}{\bm e}
\newcommand{\by}{\bm y}
\newcommand{\bF}{\bm F}
\newcommand{\bH}{\bm H}
\newcommand{\bmf}{\bm f}
\newcommand{\bh}{\bm h}
\newcommand{\bomega}{\bm \omega}
\newcommand{\sigsq}{\sigma^2}

\newcommand{\bSigmahat}{\widehat \bSigma}

\title{A flexible observed factor model with separate dynamics for the factor volatilities and their correlation matrix}

\author{Yu-Cheng Ku$^{a,b}$\footnote{Corresponding author. Tel: +61-04-32561617; fax: +61-04-93136337. {\it{E-mail}}: yku2@ncsu.edu}, Peter Bloomfield$^{a}$, Robert Kohn$^{b}$\\
\small{$^{a}$Department of Statistics, North Carolina State University, Raleigh, NC 27695, USA}\\
\small{$^{b}$School of Economics, University of New South Wales, Kensington, NSW 2052, Australia}\\}

\begin{document}
\linenumbers
\setpagewiselinenumbers
\maketitle
\bibliographystyle{asa}

\begin{abstract}
Our article considers a regression model with observed factors. The observed factors have a flexible stochastic volatility structure that has separate dynamics for the volatilities and the correlation matrix. The correlation matrix of the factors is time-varying and its evolution
is described by an inverse Wishart process. The model specifies the evolution of the observed volatilities flexibly and is
particularly attractive when the dimension of the observations is high. A Markov chain Monte Carlo algorithm is developed to estimate the model. It is straightforward to use this algorithm to obtain the predictive distributions of future observations and to carry out model selection. The model is illustrated and compared to other Wishart-type factor multivariate stochastic volatility models using various empirical data including monthly stock returns and portfolio weighted returns. The evidence suggests that our model has better predictive performance. The paper also allows the idiosyncratic errors to follow individual stochastic volatility processes in order to deal with more volatile data such as daily or weekly stock returns.

\end{abstract}

\par
\newpage
\setpagewiselinenumbers
\linenumbers

\section{Introduction}\label{S: introduction}
For the last two decades, multivariate stochastic volatility (MSV) models
have been an important class of models in financial econometrics.
Recent developments in this area focus on dimension reduction since the complexity of computation and the difficulty in model interpretation grow drastically as the dimension of the model increases. \cite{harvey:ruiz:shephard:1994} were the first to discuss a factor structure for MSV models.
The seminal work by \cite{jacquier:polson:rossi:1995} introduced Bayesian approaches to the factor MSV (FMSV) literature.
The FMSV model is also considered and discussed by \cite{pitt:shephard:1999}, \cite{chib:nardari:shephard:2006}, \cite{LopesCal}, among others.
A common feature in these FMSV models is that they impose the diagonality assumption on the factor correlation or covariance matrices,
implying that the factors are uncorrelated. However, it is often unrealistic to assume that the factors do not interact with each other,
especially when the factors are observed.

To relax the diagonality assumption, \cite{philipov:glickman:2006b} introduces a time-varying FMSV model in which the inverse factor covariance matrices are driven by Wishart processes. The model is a direct application of \cite{philipov:glickman:2006a} to the factor structure. The inverse Wishart specification introduced by Philipov and Glickman (2006a,b) has the attractive property that it can be easily incorporated into model estimation with Bayesian Markov chain Monte Carlo (MCMC) methods. Based on a similar setting, \cite{asai:mcaleer:2009} also proposes an MSV model, where the individual return series is modeled with the stochastic volatility (SV) process and the covariance process is characterized by the inverse Wishart distribution. \cite{asai:mcaleer:2009} call this type of model a ``Wishart Inverse Covariance''  (WIC) model.

When the vector of dependent variables is high-dimensional, the WIC models proposed by \cite{philipov:glickman:2006a} and \cite{asai:mcaleer:2009} have two problems. First, the computation becomes highly time-consuming as the dimension increases.
Second, the time effect among the different series is controlled by just one scalar persistence parameter, which is likely
to be too restrictive in real applications. The factor structure proposed in \cite{philipov:glickman:2006b} helps resolve the first problem. However, when it comes to more factors, say, three or more, these distinct underlying factors still have to share a common time effect controlled by a single persistence parameter; hence, the second problem remains. In order to solve the two problems
 simultaneously, we propose in our article an observed dynamic-correlation FMSV model (O-DCFMSV).

The basic model form of O-DCFMSV is similar to that of \cite{asai:mcaleer:2009},
but the structure is applied to a factor model. Consequently, compared to \cite{asai:mcaleer:2009},
the O-DCFMSV model has advantages in both model estimation and in interpretation for high-dimensional data.
Moreover, since in the O-DCFMSV model we allow different time effects on the factors through separate SV processes,
it is more flexible compared to the \cite{philipov:glickman:2006b}. To estimate the model, we develop an MCMC algorithm
that deals with all unknown parameters and latent variables
jointly, which is quite different from the partial MCMC approach used in \cite{asai:mcaleer:2009}.
Using our approach, prediction and model selection become straightforward, issues that are
not dealt with in either \cite{philipov:glickman:2006b} or \cite{asai:mcaleer:2009}. We illustrate how to implement the one-step-ahead prediction, by which we can forecast many quantities of interest, such as the return series, the return covariance matrix, the correlation matrix of the factors, and the value at risk (VaR) of a portfolio. We can also conduct model selection based on predictive performance.

To summarize, the contribution of our paper is twofold. First, it introduces a flexible factor model to the MSV literature. Secondly, the MCMC algorithm designed in this paper can be used for prediction and model selection, which significantly extends the usefulness of the WIC models in real problems.

The remainder of the paper is organized as follows. Section~\ref{S: model} presents the model and the MCMC algorithm to estimate the model.
Section~\ref{S: simulation} conducts a simulation study to illustrate the model. Section~\ref{S: empirical study}
provides two empirical examples. The O-DCFMSV model is applied to portfolio and stock return data and is compared to the model of \cite{philipov:glickman:2006b} based on the quality of
one-step-ahead predictions. Section~\ref{S : sv errors} extends the model to the case where the idiosyncratic error terms are allowed to
follow independent SV processes. Section~\ref{S: conclusion} concludes the paper.

\section{The Model}\label{S: model}
\subsection{Model Specification}\label{spec}
Suppose that at time $t$ we have $p$
asset returns, $\by_t$, and $q$ underlying observed factors, $\bmf_t$, such that
\begin{equation}
\begin{aligned}\label{eq: obs eqn}
\by_t  & = \bB \bmf_t + \be_t,
\end{aligned}
\end{equation}
where the $\{ \bmf_t, t \geq 1\}$ and $\{\be_t , t \geq 1\}$ are independent stochastic processes. The $\be_t$ are also
assumed to be independent with $\be_t \sim \text{N}_p(\be_t|\bm 0, \bOmega), \  \bOmega = \mbox{diag}(\sigsq_1,...,\sigsq_p)$, where $\text{N}_p(\bX|\bmu, \bSigma) $ is a $p$-dimensional multivariate normal density
in $\bX$ with mean $\bmu$ and covariance matrix $\bSigma$. The assumption that the conditional variance of
$\be_t$ is constant is relaxed in Section~\ref{S : sv errors} to allow
$\be_t$ to have SV dynamics. The model for the factors is as follows:
\begin{subequations} \label{eq: ft dynamics}
\begin{align}
\bmf_t & = \bm V_t^{1/2} \beps_t,\label{2a}\\
\bm V_t^{1/2} & = \mbox{diag}\left(e^{h_{t1}/2}, e^{h_{t2}/2},...,e^{h_{tq}/2}\right), q\leq p,\label{2b}\\
\bh_{t+1} & = \bm \mu + \bm \phi \circ (\bh_t - \bm \mu)+ \bm \eta_t,\label{2c}\\
h_{1i} & \sim \mbox{N} \bigg(h_{1i}\Big|\mu_i, \frac{\sigma_{\eta,i}^2}{1-\phi_i^2}\bigg), i=1,2,...,q,\label{2d}
\end{align}
\end{subequations}where $\text{N}(x|\mu, \sigma^2)$ is a univariate normal distribution in $x$ with mean $\mu$ and $\sigma^2$, and $\circ$ is the elementwise multiplication operator.
The stochastic sequences $\{\beps_t, t \geq 1\}$ and
$\{\bmeta_t , t \geq 1\}$ are independent with $\bmeta_t$ also an independent sequence and
\begin{subequations} \label{eq:  eps adn eta distns}
\begin{align}
\beps_t|\bP_t & \sim \mN_q(\beps_t| 0, \bSigma_{\eps,t}), \label{3a} \\
\bmeta_t &  \sim \mN_q(\bmeta_t| 0, \bSigma _\eta)  \ , \quad \bSigma_\eta = \mbox{diag}\left(\sigma_{\eta,1}^2,..., \sigma_{\eta,q}^2\right).\label{3b}
\end{align}
\end{subequations}

The covariance matrix $\bSigma_{\epsilon ,t}$ is a correlation matrix which
is obtained by standardizing the
$q \times q $ stochastic covariance matrix $\bP_t$ so that
\begin{align} \label{eq: stand sigma eps}
\bSigma_{\eps,t} & = (\diag \bP_t) ^{-\half} \bP_t
(\diag \bP_t) ^{-\half}.
\end{align}
The dynamics of $\bm P_t$, and hence $\bSigma_{\eps,t} $ are given by the stationary
autoregressive inverse Wishart process
\begin{align}\label{eq: dynamics Pt}
\bP_{t+1}^{-1} |k, \bP_t^{-1} & \sim \text{W}_q (\bP_{t+1}^{-1} |k, \bS_t) , \quad \bS_t = \frac{1}{k}\bP_t^{-\frac{d}{2}} \bA \bP_t^{-\frac{d}{2}},
\end{align}
where $\text{W}_q(\bm X|k, \bS)$ is a $q \times q $ Wishart density in $\bm X$ with degrees of freedom (df) $k\geq q$ and the scale matrix $\bS$. The $q \times q $ matrix $\bA$ is a symmetric positive definite matrix parameter, and $d$ is a scalar parameter that accounts for the memory of the matrix process $\{\bP_t\}$. The matrix power operation $\bP_t^{-d/2}$ is defined by a spectral decomposition. Similarly to Philipov and Glickman (2006a,b) and \cite{asai:mcaleer:2009}, we set the initial value $\bP_0$ to be $\bP_0 = \bm I_q$ for convenience.

In the WIC context, there are two different ways to define the scale matrix. \cite{asai:mcaleer:2009}
uses the specification~\eqref{eq: dynamics Pt}, while Philipov and Glickman (2006b)
uses a BEKK-type representation
\begin{equation}\label{PGspec}
\bS_{t-1} = \frac{1}{k}\bA^{\frac{1}{2}}\left(\bP_{t-1}^{-1}\right)^d (\bA^{\frac{1}{2}})^\prime,
\end{equation}
where $A^{\frac{1}{2}}$ is defined by a Cholesky decomposition such that $\bA = \bA^{\frac{1}{2}} (\bA^{\frac{1}{2}})^\prime$.
In either case, \cite{philipov:glickman:2006a} and \cite{asai:mcaleer:2009} show that $\log |\bP_{t+1}|$ is a first-order autoregression with autoregressive parameter $d$; if $d\in (-1,1)$, then this first order autoregressive process is stationary. We have also conducted simulations
that suggest that the whole process $\bm P_t$ is stationary for $d\in (-1,1)$.

Although the O-DCFMSV model has some similarities with \cite{asai:mcaleer:2009} in model specification,
the two models are in fact different in several respects. First, \cite{asai:mcaleer:2009} adopt the settings (\ref{2a}) -- (\ref{eq: dynamics Pt}) to model the return series, while in O-DCFMSV, we apply the settings to the observed factors. As a matter of fact, this is the main advantage of O-DCFMSV. When it comes to a high-dimensional environment, the estimation of the model of \cite{asai:mcaleer:2009} is extremely tedious. The reason is that the model itself is defined through a sequence of spectral decompositions or singular value decompositions, and the posterior densities of the parameters are complicated and dependent on the data dimension. The second difference is in the sampling scheme to estimate the model which we discuss below.

\subsection{Priors}
There are two set of parameters in the measurement equation \eqref{eq: obs eqn}.
For $\bOmega = \mbox{diag}\left(\sigsq_1,...,\sigsq_p\right)$, following \cite{Liesenfeld}, we assign independent inverse gamma priors for the idiosyncratic variances $\sigsq_j$. Specifically, $\sigma_j^2 \sim \mbox{IG}(\sigma_j^2|shape = \nu_0 /2, \ scale = \nu_0 s_0/2), \ j=1,...,p$. In all our analyses, we use $\nu_0=10$ and $s_0 = 0.01$. This defines a vague prior which is commonly adopted in the literature. For the loading matrix $\bB$, following \cite{jacquier:polson:rossi:1995}, we choose the prior given by
\begin{align} \label{eq: prior for B}
p(\bB|\bOmega) & \propto |\bOmega| ^{-p/2} \etr \left ( - \half \bOmega^{-1} \bB\bB^\prime\right ) ,
\end{align}
where $\etr(\bX)$ means $\exp ( \tr (\bX) ) $. This prior implies that
the columns $\bB_i$ of $\bB$ are a priori independent, each with a prior $\text{N}_p(\bB_i|\bm 0,\bOmega)$,
which is uninformative relative to the data.

The priors for the SV parameters are as follows. We adopt the default settings by \cite{kim:shephard:chib:1998}. For the mean $\mu_i$ and variance $\sigma_{\eta,i}^2, \ i=1,..,q$, we respectively assume that
$\mu_i \sim\mbox{N}\left(\mu_i\big|0, 10\right)$ and $\sigma_{\eta,i}^2 \sim \text{IG}\big(\sigma_{\eta,i}^2\big|5,0.05).$ The prior for $\phi_i$ is a shifted and scaled beta distribution. Let $\phi_i = 2\tilde{\phi_i}-1$ where $\tilde{\phi_i}\sim \text{Beta}(\phi^{(1)}, \phi^{(2)}).$ We choose $\phi^{(1)}=20$ and $\phi^{(2)}=1.5$, implying a prior mean of $2\phi^{(1)}/(\phi^{(1)}+\phi^{(2)})-1 = 0.86$.

The priors for the correlation-level parameters are chosen as follow: for $\bA$ we specify the prior $\bA^{-1} \sim \text{W}_q(\bA^{-1}|q, \ q^{-1}\bm I_q)$, which implies a prior mean of $\bm I_q$; for $d$, we choose the vague prior $d\sim \text{Unif}(d|-1,1)$. Finally, for $k$ we set $k \sim \lambda_0 e^{-\lambda_0}I_{(q,\infty)}(k)$. Note that the prior for $k$ is a truncated exponential distribution with a $rate$ parameter $\lambda_0$. Throughout the paper we set $\lambda_0 = 0.02$. This implies a prior mean of $50+q$ and a prior standard deviation of $50$, which specifies a diffuse prior.

\subsection{MCMC Estimation}
\subsubsection{Joint Distribution}
We estimate the model using the MCMC simulation method described below. Let the observed data $\bY =
\{\by_t\}:{T\times p}$, $\bF=\{\bmf_t\}:{T\times q}$, the log volatilities $\bm H= \{\bh_t\}:{T\times q} $, the normalized factors $\beps = \{\beps_t\}:{T\times p}$, and the sequence of unnormalized covariance matrices $\bP = \{\bP_t, t=1, \dots, T\}$. Let $\bomega = \{\bomega_i, i=1,...q\}$, with $\bomega_i = \{\bm \mu_i, \bm \phi_i, \bm \sigma_{\eta,i}, i=1,...,q\}$, be the parameters of the volatilities of the factors.

The joint density of $(\bY,\bF, \bH, \beps, \bP, \bB, \bOmega, \omega, \bA, d, k) $ is
\begin{align} \label{eq: joint density}
p(\bY,\bF,\bH,\beps, \bP, \bB, \bOmega, \bomega, A,d,k) & = p(\bY|\bB,\bF,\bOmega)
p(\bF|\bH,\beps) p(\bH|\bomega)p(\beps|\bP)\\
& \times p(\bP|\bA,d,k) p(\bB|\bOmega) p(\bOmega) p(\bA) p(d)p(k),\notag
\end{align}where
\begin{subequations} \label{eq: parts of like}
\begin{align}
p(\bY|\bB, \bF,\bOmega) & = \prod_{t=1}^T p(\by_t|\bmf_t,\bOmega), \label{15a}\\
p(\bF|\bH,\beps) & = \prod_{t=1}^T p(\bmf_t|\bh_t,\beps_t), \label{15b}\\
p(\bh_t|\bomega) & = p(\bh_1|\bomega) \prod_{t=2}^T p(\bh_t|\bh_{t-1},\bomega), \notag \\
p(\bh_1|\bomega) & = \prod_{i=1}^q p(h_{1i}|\bomega_i) , \quad p(\bh_t|\bh_{t-1}, \bomega) = \prod_{i=1}^q p(\bh_{ti}|\bh_{t-1,i},\bomega_i), \label{15c} \\
p(\beps|\bP, \bA) & = \prod_{t=1}^T p(\beps_t|\bP_t), \label{15d} \\
p(\bP|\bA,d,k) & = \prod_{t=1}^T p(\bP_t|\bP_{t-1},\bA,d,k). \label{15e}
\end{align}
\end{subequations}
The densities $p(\by_t|\bmf_t,\bOmega) $ in \eqref{15a} are given by Eq. \eqref{eq: obs eqn}.
The densities $p(\bmf_t|\bh_t,\beps_t)$ in \eqref{15b} are degenerate and are given by \eqref{2a}. The densities $p(h_{1i}|\bomega_i)$ in
\eqref{15c} are given by \eqref{2d}, and the densities
$p(\bh_{ti}|\bh_{t-1,i},\bomega_i)$ in \eqref{15c} are given by \eqref{2c}. The densities
$p(\beps_t|\bP_t)$ in \eqref{15d} are given by \eqref{3a} and \eqref{eq: stand sigma eps}. The densities
$p(\bP_t|\bP_{t-1},\bA,d,k)$ in \eqref{15e} are given by \eqref{eq: dynamics Pt}. The priors
$p(\bB|\bOmega)$, $p(\bOmega)$, $p(\bA)$, $p(d)$, and $p(k)$ are discussed in the previous section.

\subsubsection{Conditional Distributions}\label{Cond}
We sample from the following conditional distributions. For $\sigsq_j$, we sample from the inverse gamma distribution:
\begin{equation}\label{eq 3.10}
\begin{aligned}
p(\sigsq_j|\text{rest})  &\propto p(\sigsq_j)\cdot p(\by_j|\sigsq_j, \bB, \bF)\\
                        &\propto (\sigsq_j)^{-\frac{\nu_0+T}{2}-1}\exp\Bigg\{-\frac{1}{2\sigsq_j}\left[\nu_{0j}s_{0j} + \sum_{t=1}^T\bigg(y_{tj}-\sum_{i=1}^q b_{ji}f_{ti}\bigg)^2\right]\Bigg\},
\end{aligned}
\end{equation}
where $y_{tj}$ is the $j$th element of $\by_t$, $f_{ti}$ is the $i$th element of $\bmf_t$, and $b_{ji}$ denotes the $ij$th element of $\bB$. It follows from \eqref{eq 3.10} that the conditional density of $\sigma^2_j$ is an inverse gamma with the shape parameter $\frac{\nu_0+T}{2}$ and the scale parameter $\frac{1}{2}\left[\nu_{0j}s_{0j} + \sum_{t=1}^T\bigg(y_{tj}-\sum_{i=1}^q b_{ji}f_{ti}\bigg)^2\right]$.

The posterior density of $\bB$ is a matrix variate normal density given by:
\begin{equation}\label{eq 3.9}
\begin{aligned}p(\bB|\text{rest})& \propto p(\bB|\bOmega)\cdot p(\bY|\bB, \bOmega, \bF)\\
                                    & \propto \etr\bigg(-\frac{1}{2}\bigg\{\bOmega^{-1}\left[(\bB - \bm \mu_B)\bSigma_B^{-1}(\bB - \bm \mu_B)^\prime\right]\bigg\} \bigg),
\end{aligned}
\end{equation}
where $\bSigma_B = (\bF^\prime\bF
+ \bm I)^{-1}$ and $\bm \mu_B = \bY^\prime\bF \bSigma_B$.

We now follow \cite{kim:shephard:chib:1998} and discuss how to sample the SV parameters $\bomega$ and $\bm H$. First we transform the SV equation \eqref{2b} into a linear model by taking the logarithm:
$$
f_{ti}^* = h_{ti} + z_{ti},
$$
where $f_{ti}^* = \log(f_{ti}^2 + c)$ and $z_{ti}$ is a log $\chi^2_1$ random variable. The scalar $c$ is an ``offset'' constant that is set to be $10^{-5}$. Following \cite{kim:shephard:chib:1998}, the distribution of $f_{ti}^*$ can be approximated by a seven-component normal mixture with the component indicator variables $\bm s = \{s_{ti}\}$. Using the offset mixture integration sampler developed by \cite{kim:shephard:chib:1998}, for each $i=1,..,q$ we sample $(\bm \phi_i, \bm \sigsq_{\eta,i})$ jointly in one block marginalized over $\bm \mu_i$ and $\bm H$ and then in another block sample $(\bm \mu_i, \bm H)$ conditional on the rest in the model. To save computational cost, we do not impose the additional reweighting step introduced in \cite{kim:shephard:chib:1998}.

Given that $\bm H$ is drawn, we can then obtain $\beps_t = \bm V^{-1/2}_t \bmf_t^*$ to estimate the correlations and the correlation-level parameters. Now, since
the factors are observed and we have $\beps_t$, the estimation procedure for $\bP_t$, $\bA$, $d$, and $k$ is exactly the same as that given in \cite{asai:mcaleer:2009}. To sample from the complicated and non-conjugate univariate posterior distributions of $d$ and $k$, following \cite{asai:mcaleer:2009}, we adopt the adaptive rejection Metropolis sampling (ARMS) of \cite{gilks:best:tan:1995}. The complete MCMC procedure is given as follows:\vspace{0.3cm}
$ $\\
Step 0: Initialize $\bB, \bOmega, \bm s, \bomega, \bm H, k, d$,
and $\bA$.\\
Step 1: Sample $\bB|\text{rest}$, then sample $\sigsq_j|\text{rest}$ for $j =
1,...,p.$\\
Step 2: Sample $\bm \phi, \bm \sigma^2_{\eta}|\bF^*, \bm s$ and $\bm \mu, \bm H|\bF^*, \bm s, \bm \phi, \bm \sigma_{\eta}^2$ using the sampler of \cite{kim:shephard:chib:1998}.\\
Step 3: Obtain the standardized factors $\beps_t = \bm V^{-1/2}_t \bmf_t^*$ from the sample.\\
Step 4: Sample $\bP_t$ from $\bP_t|\text{rest}$, and then obtain $\bSigma_{\epsilon ,t}=(\diag \bP_t) ^{-\half} \bP_t
(\diag \bP_t) ^{-\half}$ for $t=1,...,T$.\\
Step 5: Sample $\bA|\text{rest}$.\\
Step 6: Sample $d|\text{rest}$ using ARMS.\\
Step 7: Sample $k|\text{rest}$ using ARMS.\\
Step 8: Go to step 1.

Looping steps 1 to 8 is a complete sweep of the MCMC sampler. It is worth noting that Steps 2 to 4 link the SV processes to the
factor correlations. To deal with this part, \cite{asai:mcaleer:2009} adopt a two-stage procedure. In the first stage they estimate the SV parameters $\bomega$ and the log-volatilities $\bm H$ in one MCMC procedure and obtain the standardized series $\epsilon_{ti} = U_{ti} f_{ti}$ with $U_{ti} = \frac{1}{M}\sum_{l=1}^M \exp\left[-\frac{1}{2}h_{ti}^{(l)}\right]$, where $x^{(l)}$ denotes the $l$th draw of the $M$ MCMC iterations. Then, in the second stage, based on the series $\beps_t = (\epsilon_{t1},...,\epsilon_{tq})$, they estimate $\{\bP_t\}$ and the correlation parameters $(\bA, d, k)$ using another MCMC procedure. Clearly, the strategy does not conduct the MCMC estimation in a joint sense, which is arguably undesirable and
improper in at least two respects. First, the method obtains the estimates the log-volatilities first and then plug in the estimates to run another separate MCMC. This manner averages out different samples of volatilities. Therefore, in the inference of the correlation-level parameters, we actually work with only one fixed set of log-volatilities and residuals $\beps$. Secondly, for the purpose of prediction,
we need to sample $\bh_{t+1}^{(l)}$ and then obtain $\bSigma_{\epsilon ,t}^{(l)}$, for $l=1,...,M$. The plug-in method cannot be applied in this case. Unlike the two-stage scheme of \cite{asai:mcaleer:2009}, our algorithm makes draws for $\bomega$ and $\bm H$ and then directly obtains the standardized series for the correlation parameters in each single iteration. In this way, we conduct estimation jointly with a full MCMC procedure, and the prediction can be performed directly using the usual MCMC methods.

\section{Simulation Study}\label{S: simulation}
\subsection{Simulated Data Example}\label{simDGP}
In this subsection, we use simulated data to illustrate how O-DCFMSV works.
Note that the illustration is based on a single run since at the very beginning we wish to present the result visually.
A complete simulation study based on multiple replications is provided later. We set $p=10$ observed series and $q=2$ factors with a sample size $T=1,000$. The true data generating process (DGP) is described by:

\begin{flushleft}(i) Measurement equation: \vspace{-0.2cm}\end{flushleft}
$$
\begin{aligned}
\bB & = \left(\begin{array}{rrrrrrrrrr}
1.00 & 0.30 & -0.05 & 0.99 & 0.99 & -0.10 & 0.00 & 0.56 & 0.00 & 0.00\\
0.00 & 1.00 & 0.34 & 0.00 & 0.00 & 0.95 & 0.95 & 0.00 & 0.00 & 0.30\\
\end{array}\right)^\prime,\\
\bOmega & = \mbox{diag}(0.05,\ 0.1,\ 0.13,\ 0.24,\ 0.35,\ 0.35,\ 0.24,\ 0.13,\ 0.1,\ 0.05).
\end{aligned}
$$

\begin{flushleft}(ii) SV structures: \vspace{-0.5cm} \end{flushleft}
$$
\begin{aligned}
h_{1,t+1} &= \mu_1 + 0.95( h_{1t}-\mu_1) + \eta_{1t}, \ \mu_1 = -0.2, \\
h_{2,t+1} &= \mu_2 + 0.98 ( h_{2t}-\mu_2) + \eta_{2t}, \ \mu_2 = -0.5, \vspace{0.3cm}\\
\left(\begin{array}{c}\eta_{1t}\\
\eta_{2t}\end{array}\right) & \sim \mbox{N}\left(\left[\begin{array}{c}0\\
0\end{array}\right], \left[\begin{array}{cc}0.1^2 & 0\\
0 & 0.27^2\end{array}\right]\right)\vspace{0.5cm}.
\end{aligned}
$$

\begin{flushleft}(iii) Factor correlation level: \vspace{-0.2cm} \end{flushleft}
$$
\bA = \left(\begin{array}{cc}1 &0.05\\
0.05 & 1\end{array}\right)^{-1} = \left(\begin{array}{cc}1.003 & -0.050\\
-0.050 & 1.003\end{array}\right), k = 25, \ d = 0.8,
$$ From (iii) and the initial value $\bP_0 = \bm I_q$ we can simulate a sequence of covariance matrices $\{\bP_t\}$, from which we can obtain the correlation matrices $\bSigma_{\epsilon, t}$ using Eq. \eqref{eq: stand sigma eps}. Given $\{\bSigma_{\epsilon, t}\}$ together with (ii), we can generate two hidden systematic factors with time-varying correlation $\rho_t = [\bSigma_{\epsilon, t}]_{2,1}$. Then, given the factors we can generate ten observed series $\bY$ with the setting (i).

The MCMC study is conducted with 20,000 iterations, where the first $L=10,000$ draws are taken as burn-ins and the remaining $M=10,000$ are preserved. The program for the estimation is coded in OX by \cite{doornik:2007}. Table~\ref{table1} summarizes the estimation results. We output the posterior means and the 95\% intervals based on the $M$ draws. The posterior mean is calculated by averaging the MCMC draws. The 95\% credible interval is constructed using the (2.5\%, 97.5\%) percentiles of the simulated draws. We can see that, out of the $pq + p + 3q + q(q+1)/2 + 2 = 41$ parameters, there is only one, $b_{72}$, not covered by the 95\% credible interval.

\begin{figure}
\begin{center}
\includegraphics[height=3in]{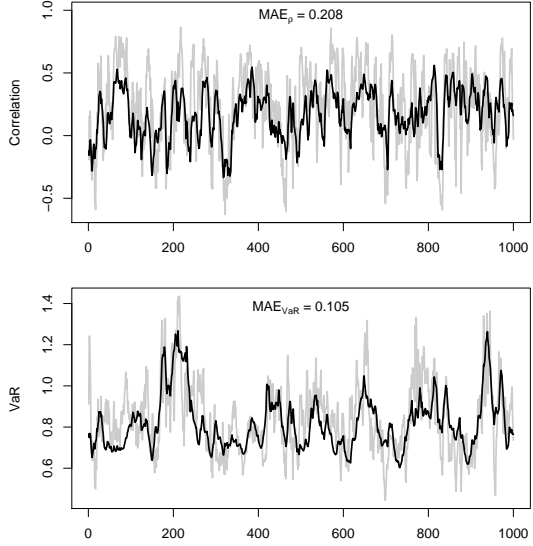}
\caption{Factor correlations and VaR estimates. The top panel shows the true correlation process $\rho_t$ (grey line) and its posterior mean $\hat{\rho}_t$ (solid black line). The bottom panel shows $\text{VaR}^0_t$ (grey line) and its posterior mean $\text{VaR}^{est}_t$ (solid black line).}\label{fig1.1}
\end{center}
\end{figure}

In O-DCFMSV, one of the primary interests is in capturing the time-varying factor correlation.
The factor correlation provides very useful information since it can reflect the market condition
as we will see later.
The top panel of Figure \ref{fig1.1} displays the correlation fits. The smoothed estimate at $t$ is calculated by the posterior mean $\hat{\rho}_t =M^{-1} \sum_{l=1}^M [\bSigma_{\epsilon ,t}]_{21}^{(l)}$,
where $\bSigma_{\epsilon ,t}^{(l)}$ is the $l$th draw of the preserved MCMC iterations based on smoothing.
Note that we draw $\bSigma_{\epsilon ,t}^{(l)}$ using Step 4 in the algorithm detailed in Section~\ref{Cond}.
The grey line is the true correlation and the black line represents the fits. We can observe that, though the fitted result appears smoother than
the true values, the model in general well captures the pattern of the dynamic correlations, both the movements and the average level. To measure the performance, following \cite{asai:mcaleer:2009}, we calculate two performance measures, both of which are based upon the mean absolute error (MAE) of the smoothing estimates. The first is $\mbox{MAE}_\rho \equiv \frac{1}{T}\sum_t|\hat{\rho}_t-\rho_t|$, which measures the quality of the correlation estimates. We obtain the $\text{MAE}_\rho  =  0.208$. This suggests a satisfactory result, as we can see that the correlation varies
in the (wide) range $(-0.6, 0.9)$.

The second measure is used for evaluating the VaR estimates. This measure is meaningful to the O-DCFMSV model since one important application of the asset-return factor model is to obtain the VaR estimates of the portfolios through factor structures. Suppose that we have a vector of portfolio asset weights $\bm w$. According to \cite{barbieri:2009} and \cite{chib:nardari:shephard:2006},
under the assumptions of normality and a zero mean, the 5\% VaR of the simulated portfolio at time $t$ is estimated by
$1.645\cdot \hat{\sigma}_{P_t},$ where $\hat{\sigma}_{P_t}$ denotes the posterior
mean of the portfolio standard deviation ${\sigma}_{P_t} = \left[\bm w^\prime (\bB\bm V_t^{1/2}\bSigma_{\epsilon, t}\bm V_t^{1/2}\bB^\prime + \bOmega)\bm w\right]^{\frac{1}{2}}$. (In fact this is the 95\% quantile of the predictive density when the mean is zero,
but by symmetry it is the negative of the 5\% quantile).
Suppose that our asset holdings are equally weighted, which means that $\bm w = \frac{1}{10} \bm 1$.
The true and estimated VaR are shown in the bottom panel of Figure \ref{fig1.1}, where the grey line represents the true VaR and the black line is the estimate. It is readily seen that both the movement and the magnitude are nicely captured. The MAE measure is obtained by $\mbox{MAE}_{\mbox{VaR}} \equiv \frac{1}{T}\sum_t|\mbox{VaR}^{est}_t-\mbox{VaR}^0_t|$, where $\mbox{VaR}^{est}_t = 1.645\cdot\hat{\sigma}_{P_t}$ and $\mbox{VaR}^0_t = 1.645\cdot {\sigma}_{P_t}$. We have the result $\mbox{MAE}_{\mbox{VaR}} = 0.105$, which is quite satisfactory, for the true VaR is fluctuating in a range from 0.5 to 1.5.

\begin{spacing}{1.6}
\begin{table}[ht]
\caption{MCMC results for the simulated data. TRUE is the true value of the parameter. MEAN is the posterior mean. (LOWER, UPPER) are the boundaries of the 95\% interval.}\label{table1}
\begin{center}
\begin{tabular}{rrrrr}
  \hline
 & TRUE & MEAN & LOWER & UPPER \\
  \hline
  $B_{11}$          & 1.000     & 0.989    & 0.975     & 1.003 \\
  $B_{21}$          & 0.300     & 0.303    & 0.286     & 0.321 \\
  $B_{31}$          & -0.050    & -0.044   & -0.064    & -0.024\\
  $B_{41}$          & 0.990     & 0.969    & 0.946     & 0.993 \\
  $B_{51}$          & 0.990     & 1.009    & 0.987     & 1.031 \\
  $B_{61}$          & -0.100    & -0.080   & -0.107    & -0.053\\
  $B_{71}$          & 0.000     & 0.000    & -0.031    & 0.030 \\
  $B_{81}$          & 0.560     & 0.569    & 0.532     & 0.605 \\
  $B_{91}$          & 0.000     & -0.016   & -0.052    & 0.021 \\
  $B_{10,1}$        & 0.000     & 0.007    & -0.037    & 0.051 \\
  $B_{12}$          & 0.000     & 0.012    & -0.024    & 0.050 \\
  $B_{22}$          & 1.000     & 0.995    & 0.950     & 1.041 \\
  $B_{32}$          & 0.340     & 0.328    & 0.296     & 0.359 \\
  $B_{42}$          & 0.000     & -0.008   & -0.047    & 0.031 \\
  $B_{52}$          & 0.000     & -0.020   & -0.043    & 0.003 \\
  $B_{62}$          & 0.950     & 0.937    & 0.908     & 0.965 \\
  $B_{72}$          & 0.950     & 0.918    & 0.899     & 0.938 \\
  $B_{82}$          & 0.000     & -0.006   & -0.030    & 0.019 \\
  $B_{92}$          & 0.000     & -0.008   & -0.022    & 0.006 \\
  $B_{10,2}$        & 0.300     & 0.298    & 0.281     & 0.315 \\
  $\sigma_1^2$      & 0.050     & 0.052    & 0.048     & 0.057 \\
  $\sigma_2^2$      & 0.100     & 0.099    & 0.091     & 0.108 \\
  $\sigma_3^2$      & 0.130     & 0.124    & 0.113     & 0.135 \\
  $\sigma_4^2$      & 0.240     & 0.230    & 0.211     & 0.251 \\
  $\sigma_5^2$      & 0.350     & 0.333    & 0.304     & 0.362 \\
  $\sigma_6^2$      & 0.350     & 0.345    & 0.316     & 0.377 \\
  $\sigma_7^2$      & 0.240     & 0.253    & 0.232     & 0.276 \\
  $\sigma_8^2$      & 0.130     & 0.135    & 0.124     & 0.147 \\
  $\sigma_9^2$      & 0.100     & 0.101    & 0.092     & 0.110 \\
  $\sigma_{10}^2$   & 0.050     & 0.049    & 0.045     & 0.054 \\
  $\mu_1$           & -0.200    & -0.092   & -0.280    & 0.087 \\
  $\mu_2$           & -0.500    & -0.859   & -1.560    & -0.141\\
  $\phi_1$          & 0.950     & 0.945    & 0.878     & 0.981 \\
  $\phi_2$          & 0.980     & 0.977    & 0.960     & 0.991 \\
  $\sigma_{\eta,1}$ & 0.100     & 0.125    & 0.076     & 0.211 \\
  $\sigma_{\eta,2}$ & 0.270     & 0.222    & 0.168     & 0.285 \\
  $a_{11}$          & 1.003     & 0.934    & 0.820     & 1.034 \\
  $a_{12}$          & -0.050    & -0.042   & -0.080    & -0.018\\
  $a_{22}$          & 1.003     & 0.970    & 0.793     & 1.211 \\
  $d$               & 0.800     & 0.756    & 0.587     & 0.856 \\
  $k$               & 25.000    & 20.797   & 10.406    & 35.057\\
   \hline
\end{tabular}
\end{center}
\end{table}
\end{spacing}

\clearpage

\subsection{Model Comparison}
To complete the illustration, we finish the simulation study by comparing O-DCFMSV with the benchmark, the model of Philipov and Glickman (2006b, hereafter PG), as it is also a Wishart FMSV model with dynamic factor correlations. The model specification is:
$$
\begin{aligned}
\by_t|\bB, \bmf_t, \bm\Omega &\sim \mbox{N}_p(\bB \bmf_t, \bOmega),\\
\bmf_t|\bP_t &\sim \mbox{N}_q(\bm 0, \bP_t),\\
\bP_t^{-1}|\bP_{t-1}^{-1}, \bS_{t-1} &\sim W_q(\bP_{t-1}^{-1}| k, \bS_{t-1}),
\end{aligned}
$$
where the matrix $\bP_t$ is a factor covariance matrix, the meaning of the matrix $\bA$ and the scalar parameters $d$ and $k$ are the same as those in O-DCFMSV. Here we define the scale matrix as $\bS_t = \frac{1}{k}\bP_t^{-\frac{d}{2}} \bA \bP_t^{-\frac{d}{2}}$, which is the form of \eqref{eq: dynamics Pt}. It should be noted that, as mentioned in Section~\ref{spec}, \cite{philipov:glickman:2006b} use the BEKK-type specification (\ref{PGspec}) for $\bS_t$, however, in order to remove the effect caused by different parameterizations, we adopt the setting \eqref{eq: dynamics Pt} instead of \eqref{PGspec} for the competing model. \cite{asai:mcaleer:2009} point out that it is possible to use either one or the other as an alternative.
The matrix power $\bP_t^{-\frac{d}{2}}$ is calculated by the spectral decomposition.

We take (i), (ii), and (iii) used in last section as the true data generation process (DGP) for O-DCFMSV.
For the true DGP of PG's model, we drop (ii) and use only (i) and (iii), since the model does not assume SV structures on the factors.
Two datasets with different DGPs are generated and fitted with both models. To evaluate the performance, we calculate the Kullback-Leibler (KL) divergence as a measure of how far away the distribution given the estimated covariance is from that given the truth.
Let $\bSigma_{t}^0$ and $\hat{\bSigma}_{t}$ be the true and estimated covariance matrices, respectively. Let $p^0_t = p(\by_t|\bSigma_{t}^0)$ denote the density of $\by_t$ given true covariance matrix of $\by_t$; also, let $p^{\text{est}}_t = p(\by_t|\hat{\bSigma}_{t})$ be the density of $\by_t$ with the estimated covariance matrix plugged in instead of the true covariance matrix. Under normality, the KL divergence between $p_t^0$ and
$p^{\text{est}}_t $ is
$$
\begin{aligned}
\text{KL}\left(p^0_t\big|\big|p^{\text{est}}_t\right) & =
\int p^0_t(y)\log\frac{p^0_t(y)}{p^{est}_t(y)}dy\\
& = -\frac{p}{2} +\frac{1}{2}\text{tr}\left(\hat{\bSigma}_{t}^{-1}\bSigma_{t}^0\right) - \frac{1}{2}\log\big|\bSigma_{t}^0 \big| + \frac{1}{2}\log\big|\hat{\bSigma}_{t} \big|.
\end{aligned}
$$ In each replication we record the mean KL divergence (MKL), which is defined by $\frac{1}{T}\sum_t \text{KL}\left(p^0_t\big|\big|p^{\text{est}}_t\right)$ as a summary of the KL divergence over every $t$.

Since we have two true DGPs and two models, there are four combinations, which, by the DGP-model order, are referred as O-O, O-PG, PG-O, and PG-PG. In each combination we conduct the simulation with 40 replications and record their MKL. In each replication we calculate the {\it{differenced}} measure by subtracting the value of the true model from that of the wrong model, i.e. [O-PG minus O-O] and [PG-O minus PG-PG]. The differenced measure is denoted by $\Delta \text{MKL}$. We report the sample mean of $\Delta \text{MKL}$ from the 40 replications and the standard error as the final summarized output. Table~\ref{table2} summarizes the comparison results. We can see that, for both DGPs, the differenced values of [wrong minus true] are both significantly positive, which indicate that the true models win. However, we can find that the mean $\Delta \text{MKL}$ appears to be larger in the case when the true DGP is O-DCFMSV. This has an important implication that, given both the models are misspecified, PG's model is more distant from the truth than O-DCFMSV; in other words, the KL loss is greater when PG is used and O-DCFMSV is correct, than vice versa.

\begin{table}
\caption{Results of $\Delta \text{MKL}$ for the comparison of O-DCFMSV with PG's model using simulated data. The difference is defined by the false model minus the true model. Std Err is the standard error of the mean.}\label{table2}
\begin{center}
\begin{tabular}{lrr}
\hline
True DGP & Mean $\Delta \text{MKL}$ & Std Err\\
\hline
O-DCFMSV  & 0.012  & 0.002\\
PG        & 0.007  & 0.002\\
\hline
\end{tabular}
\end{center}
\end{table}

\section{Empirical Study} \label{S: empirical study}
\subsection{Factor Correlation}\label{FactCorrDemo}
\begin{figure}
\begin{center}
\includegraphics[height=3in]{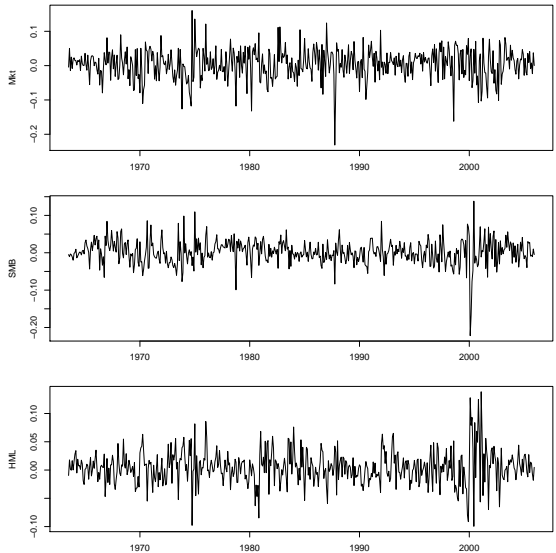}
\caption{Time-series plot of the rescaled F-F factors, Jul 1963 - Dec 2005.}\label{factplot}
\end{center}
\end{figure}

This section uses two empirical examples to illustrate the applications of the O-DCFMSV model.
We use three monthly Fama-French (F-F) factors obtained from Dr. Kenneth French's data library. The factors are the market excess return (Mkt), the Small-Minus-Big factor (SMB) and the High-Minus-Low factor (HML). All the factors are rescaled to (-1,1) by multiplying by 0.01.
The observation period is from July 1963 to December 2005 with a total sample size of 510.
Figure~\ref{factplot} shows the time series plot of the three rescaled F-F factors. We can see that, the behavior of the volatility is quite different among different factors. For example, during the 1980's and the 90's, the volatility is small for SMB, whereas it appears to be large for Mkt and HML.
In addition, we can observe that a cluster occurs around late 1990's to early 2000's in all three factors, but the magnitude of the volatility is noticeably larger in SMB and HML than in Mkt. Accordingly, we need to allow the factor volatilities to have separate dynamics in order to reflect such facts.

Prior to the examples, we first display the time series plot of the ``true'' factor correlations to show that the dynamic factor correlation is a reasonable setting. The ``true'' correlation at time $t$ is $\rho_{t-r\ :\ t+r}$, which we take simply as the empirical correlation calculated from the data within the window $[t-r,t+r]$. For example, if we choose $r=3$, then the correlation of (Mkt,SMB) at January 1980 is calculated by the empirical correlation of (Mkt,SMB) from October 1979 to April 1980, which approximately represents a half-year correlation. Here we choose $r=6$ for calculating the 1-year correlation, $r=12$ for the 2-year, and $r=18$, the 3-year. By rolling the windows, we obtain the ``true'' correlations over time.

\begin{figure}[ht]
\begin{center}
\includegraphics[height=4in]{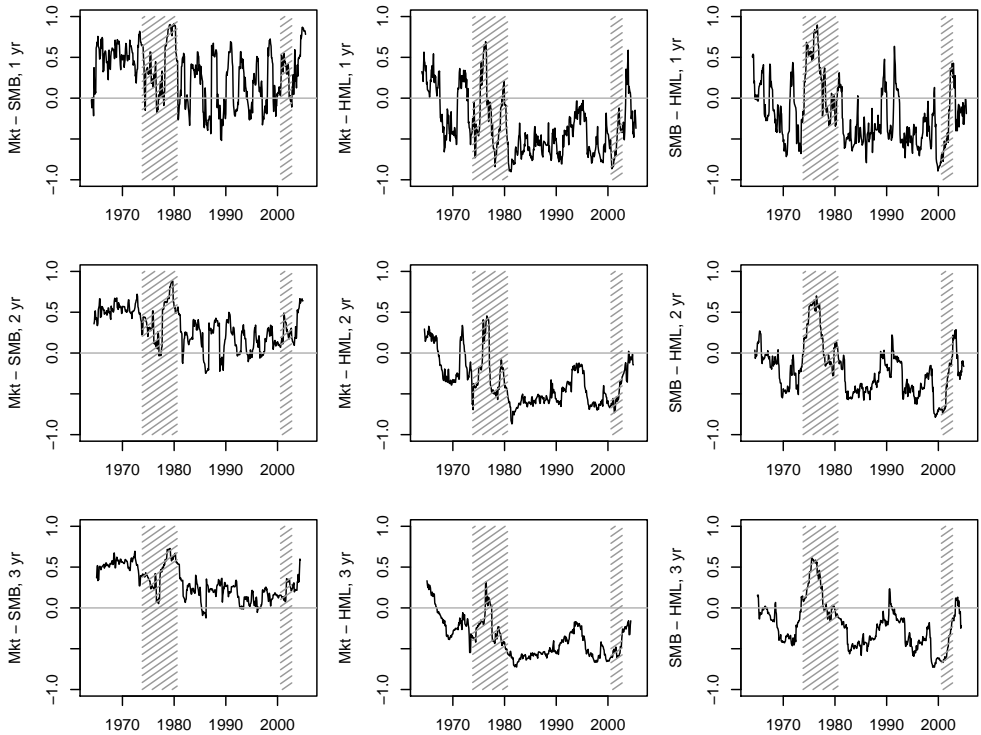}
\caption{``True'' correlations of the F-F factors Mkt-RF, SMB, and HML, Jul 1963 - Dec 2005. The shaded areas account for the events that have great economic impacts, which respectively are the first and the second oil crisis, and the Dot-com bubble burst.}\label{fig4.1}
\end{center}
\end{figure}

Figure~\ref{fig4.1} shows the 1-year, 2-year, and 3-year pairwise correlations of the three factors. It is obvious that some pairs have quite large correlations during certain periods. The shaded areas account for the events having great economic impacts, which respectively are the first and  second oil crisises, and the bursting of Dot-com bubble. Obviously, we can see that the factor correlations are changing over time. In particular, we can observe a common pattern: the correlations climb to a higher level or local peaks during these turbulent periods, while in the ``calm'' periods such as 1980's to 90's, the correlations decline to a relatively low level. This represents the well-known ``correlation breakdown'' phenomenon that has long been recognized in empirical data, referring to the pattern that the correlation during ordinary and stressful market conditions differ substantially. See \cite{rey:2000} for a detailed discussion. The correlation breakdown implied in Figure~\ref{fig4.1} suggests that time-varying factor correlation should be considered in the modeling. We would also like to point out that the main advantage of using O-DCFMSV over the empirical rolling-window method to estimate factor correlations is that we can calculate the credible intervals, by which we can obtain the significance relative to a specific critical level.

\subsection{Example 1 -- Portfolio Return Data}\label{monthlycase}
\begin{figure}[ht]
\begin{center}
\includegraphics[height=4.2in]{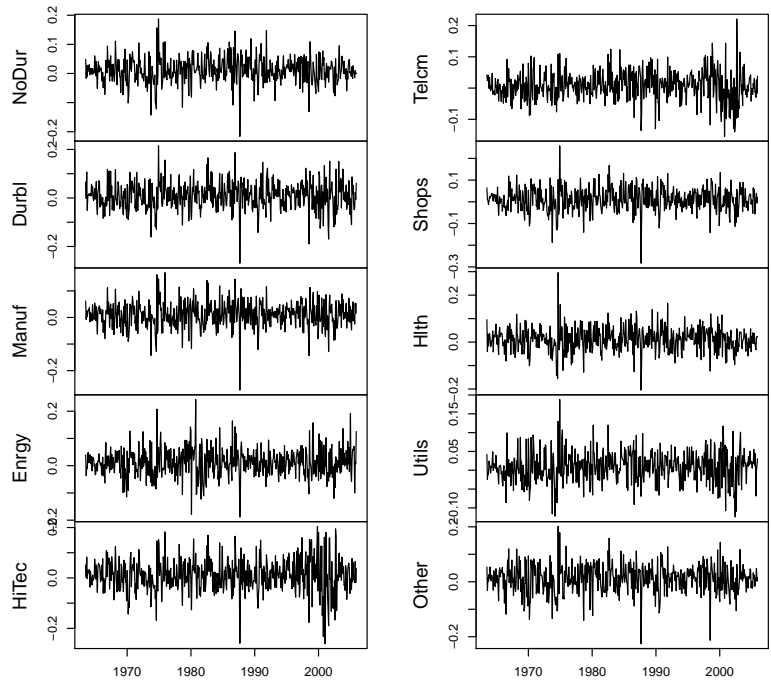}
\caption{Time-series plot of the ten portfolios, Jul 1963 - Dec 2005.}\label{fig4.2}\vspace{1.2cm}
\end{center}
\end{figure}

\begin{figure}[ht]
\begin{center}
\includegraphics[height=4.2in]{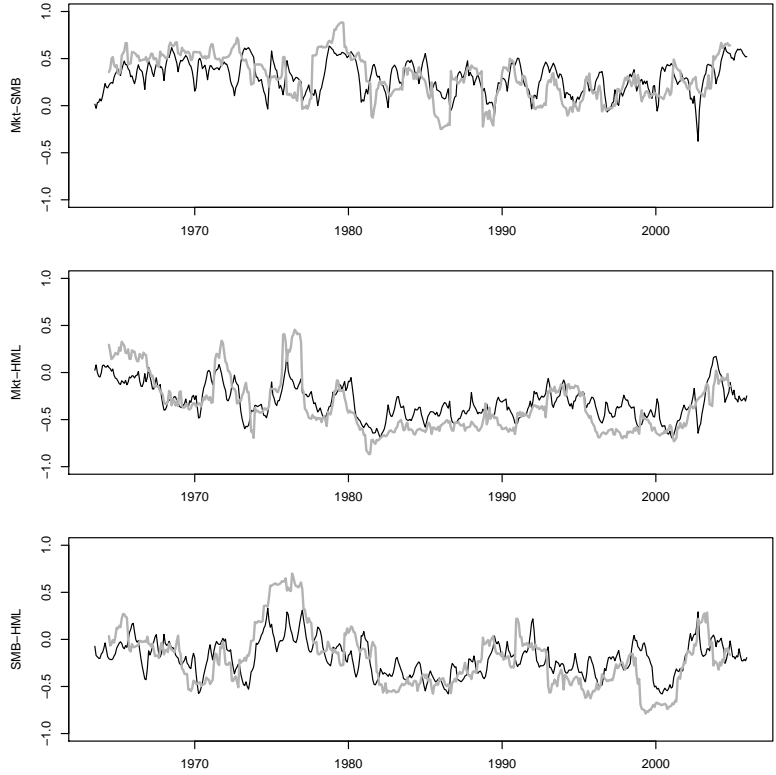}
\caption{Estimated and ``true'' correlations of the factors for the portfolio data. The black lines are the fitted values and the grey lines are the ``true'' values.}\label{fig4.3}
\end{center}
\end{figure}

In this first example we use the three F-F factors demonstrated in last section as the covariates.
The return series $\bY$ are the monthly average value weighted returns for 10 industry portfolios obtained from Dr. Kenneth French's data library.
The 10 portfolios are: NoDur, Durbl, Manuf, Enrgy, HiTec, Telcm, Shops, Hlth, Utils, and Other.
A detailed description for these portfolios can be found in the data library. Again, we first convert the data to a (-1,1) scale by multiplying by 0.01. Figure~\ref{fig4.2} shows the time series plot of the data. We can observe some common clusters occurring in the mid 1970's and the early 2000's, which suggests that the factor SV structure can be useful. Figure~\ref{fig4.3} shows the fitted factor correlations. The black lines are the fitted values and the grey lines represent the 2-year ``true'' factor correlations presented in Section~\ref{FactCorrDemo}. We see that the estimates are smoother than the ``true'' values, but in general, the movements and the magnitude of the correlations are properly captured.

We now compare the O-DCFMSV model with PG's model using several performance measures.
The first two measures are based on the one-step-ahead predictive ability for the return covariance matrix.
Notice that in this example we have 10 return series and in each month there are more than 10 transaction days.
Thus, we can use daily returns to construct a nonsingular empirical covariance matrix as a proxy for the ``true'' covariance.
Given the ``true'' covariance matrices, we can therefore compute $\mbox{MAE}_{\mbox{VaR}}$ for the equally-weighted portfolio as we do in Section~\ref{S: simulation}. The empirical covariance matrix at month $t$, denoted by $\bSigma_{t}$, is simply the sample covariance matrix constructed from the daily observations within that month with an adjustment factor of $n_t/(n_t-1)$, where $n_t$ is the number of transaction days within month $t$. The one-step-ahead predictor for the return covariance matrix given a model can be obtained by the conditional covariance:
$$\bSigma_{t+1}^{\mathcal{M}} \equiv \mbox{Var}\left(\by_{t+1}|\mathcal{F}_{t}; \mathcal{M}\right) = \bB \mbox{Var}\left( \bmf_{t+1} |\mathcal{F}_{t}; \mathcal{M}\right)\bB^\prime + \bOmega,$$ where $\mathcal{F}_{t} = \{\by_1,...,\by_t\}$ is the set of observations collected up to time $t$, and $\mathcal{M}$ denotes the model, either PG or O-DCFMSV. In the implementation, for each period $t+1$ we rerun the MCMC procedure to obtain the one-step-ahead covariance matrix. Let $\bSigma_{t+1}^{PG}$ be the one-step-ahead predictive covariance matrix of $\by_{t+1}$, which is estimated by
$$\bSigmahat_{t+1}^{PG} \approx \frac{1}{M}\sum_{l=1}^M \left[\bB^{(l)} \bm P_{t+1}^{(l)} {\bB^{(l)}}^\prime +\bOmega^{(l)}\right],$$ where $M$ is the number of preserved MCMC iterations and $\bm P_{t+1}^{(l)} \sim \bm P_{t+1}\big|\text{rest}; \mathcal{M}_{PG}$.

For the O-DCFMSV model, the factor follows:
$$\bmf_{t+1} | \bm V_{t+1}, \bSigma_{\epsilon, t+1} \sim
\mbox{N}_q\Big(\bm 0, \bm V_{t+1}^{1/2}\bSigma_{\epsilon, t+1}\bm
V_{t+1}^{1/2}\Big).
$$
Define $\bm R_{t+1} = \bm V_{t+1}^{1/2}\bSigma_{\epsilon, t+1}\bm
V_{t+1}^{1/2}$, where $\bm V_{t+1} = \diag(V_{t+1,1},...,V_{t+1,q})$.
From \eqref{2c}, we have that
\begin{subequations}\label{eq: exp h volat}
\begin{align}
V_{t+1,i}|h_{ti}, \bomega_i &  = \exp(h_{t+1,i}) |h_{ti}, \quad \bomega_i  \sim \text{logN}(V_{t+1,i}| \lambda_{ti},\sigma_{\eta,i}^2), \quad \text{where} \label{20a}\\
\lambda_{ti} & = \phi_i h_{ti} + (1-\phi_i)\mu_i \notag,  \\
V_{t+1,i}^\half |h_{ti}, \bomega_i & \sim \text{logN} (V_{t+1,i}^\half |\lambda_{ti}/2, \sigma_{\eta,i}^2/4), \label{20b}
\end{align}
\end{subequations}
and $\text{logN}(x|a,b)$ is a log normal density in $x$ where $\log x \sim \text{N}(\log x|a,b)$.
The correlation matrix $\bSigma_{\eps,t+1}$ is obtained by \eqref{eq: stand sigma eps}. Then, in the $l$th MCMC iteration we can calculate $\bm R_{t+1}^{(l)}$. Let $\bSigma_{t+1}^O$ be the one step ahead predictive variance of $\by_{t+1}$ under the O-DCFMSV model. Given $\bm R_{t+1}^{(l)}$, we have the approximation
$$\bSigmahat_{t+1}^{O} \approx
\frac{1}{M}\sum_{l=1}^M \left({\bB}^{(l)} {\bm R_{t+1}}^{(l)} {\bB^{(l)}}^\prime +\bOmega^{(l)}\right).
$$

To evaluate the predictive accuracy for the 5\% VaR predictions for the equally-weighted portfolio, we calculate
$$\mbox{MAE}_{\text{VaR}} \equiv \frac{1}{N}\sum_t\Big|\mbox{VaR}^{est}_{t+1}-\mbox{VaR}^0_{t+1}\Big|,$$ where $N$ is the number of forecast periods and the quantity $\text{VaR}^0_{t+1} = 1.645 \cdot \left(\bm w' \bSigma_{t+1} \bm w\right)^{1/2}$ with $\bm w = p^{-1}\bm 1$ being the weight vector. We calculate the estimate
$\text{VaR}^{est}_{t+1}$ using
\begin{align*}
\text{VaR}^{est}_{t+1} & = 1.645 \times \left ( M^{-1}
 \sum_{l=1}^M \left [\bm w' \bSigmahat_{t+1}^{\mathcal{M}} \bm w \right ]^{(l)}
\right) ^\half \ .
\end{align*}
In addition to the $\mbox{MAE}_{\text{VaR}}$ measure, some authors suggest calculating the difference between the ``true'' and predicted covariance matrices in an elementwise sense. Following \cite{Ledoit:Santa-Clara:Wolf:2003}, we calculate the root-mean-square error (RMSE) based on the Frobenius norm (FN):
$$
\text{FN} = \frac{1}{N}\sum_t \big|\big|\bSigma_{t+1}-\bSigmahat_{t+1}^{\mathcal{M}}\big|\big| = \frac{1}{N}\sum_t\left[\sum_{i,j}([\bSigma_{t+1}]_{ij} - [\bSigmahat_{t+1}^{\mathcal{M}}]_{ij})^2\right]^{1/2}.
$$
Because both $\text{MAE}_{\text{VaR}}$ and FN measure deviations from  ``true'' values, a smaller value indicates a better model. Here we calculate the ratio of PG to O-DCFMSV for these measures so that we can compare the deviations. We output the mean values of the ratios over all the prediction periods as the summarized results.

Besides using the empirical covariance error-based measures, we also evaluate model performance in terms of the predictive quality for the return series. To do this, following \cite{geweke:amisano:2010}, we first obtain the one-step-ahead log predictive score (LPS) and then calculate the cumulative log predictive Bayes factor. The one-step-ahead LPS evaluated at $\by_{t+1}$ under the specific model $\mathcal{M}$ is given by
$$
\text{LPS}(\by_{t+1}|\mathcal{F}_{t}; \mathcal{M})  = \log p(\by_{t+1}|\mathcal{F}_{t}; \mathcal{M}),
$$
where the predictive density $p(\by_{t+1}|\mathcal{F}_{t}; \mathcal{M})$ is calculated by
$$
\begin{aligned}
p(\by_{t+1}|\mathcal{F}_{t}; \mathcal{M})&= \int p(\by_{t+1}|\mathcal{F}_{t};\bm \theta_{\mathcal{M}})p(\bm \theta_{\mathcal{M}}|\mathcal{F}_{t})d\bm \theta_{\mathcal{M}}\\
& \approx \frac{1}{M}\sum_{l=1}^M p\left(\by_{t+1}\big|\bm x_{t+1}^{(l)}, \bm \theta^{(l)}_{\mathcal{M}}\right)\\
& = \frac{1}{M}\sum_{l=1}^M \text{N}_p\left(\by_{t+1}\big|\bB^{(l)} \bmf_{t+1}^{(l)}, \bOmega^{(l)}; \mathcal{M}\right),
\end{aligned}
$$ where $\theta_{\mathcal{M}}$ is the set of parameters for the model $\mathcal{M}$ and $\bm x_{t+1}$ is the latent state vector.
Then we can calculate the cumulative log predictive Bayes factor of Model 1 against Model~0, which is defined by
$$
\begin{aligned}
\log (B_{1,0}) &= \sum_{t} \log p(\by_t| \mathcal{F}_{t-1}; \mathcal{M}_1) - \sum_{t} \log p(\by_t| \mathcal{F}_{t-1}; \mathcal{M}_0)\\
& = \sum_{t}\Big[ \text{LPS}(\by_{t}|\mathcal{F}_{t-1}; \mathcal{M}_1) - \text{LPS}(\by_{t}|\mathcal{F}_{t-1}; \mathcal{M}_0) \Big].
\end{aligned}
$$ In addition, we calculate the LPS for the equally-weighted portfolio $\bm w^\prime \by_{t+1}$, say, LPS-EW:
$$
\begin{aligned}
\text{LPS-EW}(\bm w' \by_{t+1}|\mathcal{F}_{t}, \mathcal{M}) &= \log P(\bm w' \by_{t+1}|\mathcal{F}_{t}, \mathcal{M})\\
 &\approx \log \left[\frac{1}{M}\sum_{l=1}^M \text{N}\left(\bm w' \by_{t+1}\big|\bm w' \bB^{(l)} \bmf_{t+1}^{(l)}, \bm w' \bOmega^{(l)} \bm w \right)\right].
\end{aligned}
$$ The reason to calculate LPS-EW is that if the model performs better in this measure, then we have evidence to believe that the model should also be better in forecasting the VaR for an equally-weighted portfolio. In this sense, we can regard LPS-EW as an alternative to $\text{MAE}_{\text{VaR}}$. Similar to LPS, we then calculate the cumulative log predictive Bayes factor of Model 1 against Model 0 for the equally-weighted portfolio:
$$
\log (B^{EW}_{1,0}) = \sum_{t}\Big[ \text{LPS-EW}(\by_{t}|\mathcal{F}_{t-1}; \mathcal{M}_1) - \text{LPS-EW}(\by_{t}|\mathcal{F}_{t-1}; \mathcal{M}_0) \Big].
$$ The cumulative log predictive Bayes factor has a simple criterion for checking statistical significance. According to \cite{geweke:amisano:2010}, the evaluation is conducted via the log scoring rule described in \cite{gneiting:raftery:2007}. The detailed log scoring rule is given in \cite{kass:raftery:1995}, of which we use the following criterion: if $\log (B_{1,0})<0$, the evidence is in favor of Model 0; if $\log (B_{1,0})\in [0, 1)$, the evidence is not worth more than a bare mention; if $\log (B_{1,0})\in [1,3)$, the evidence is positive in favor of Model 1; if $\log (B_{1,0})\in [3,5)$, the evidence is strongly in favor of Model 1; if $\log (B_{1,0})>5$, we have {\it{very strong}} evidence in favor of Model 1.

We use a three-year out-of-sample prediction period, from January 2006 to December 2008, with a total length $N=36$. This time frame covers two market conditions: before 2007, when the market is relatively calm, and afterwards when the market is relatively volatile due to the subprime crisis. Therefore, we can compare model performance across different market conditions. Here, the one-step-ahead prediction is conducted on a rolling basis, i.e., if we use observations $\by_1,...,\by_T$ to forecast $\by_{T+1}$, then in next period, $\by_{T+1}$ is included as a sample for the prediction of $\by_{T+2}$. Table~\ref{table3} (a) summarizes the results of the comparison using the empirical covariance error-based measures. $\text{R-MAE}_{\text{VaR}}$ and $\mbox{R-FN}$ denote the ratio of PG to O-DCFMSV for $\text{MAE}_{\text{VaR}}$ and $\mbox{FN}$, respectively. We see that the mean ratio for $\text{MAE}_{\text{VaR}}$ is 1.22, suggesting that the deviance of PG is $22\%$ larger than that of O-DCFMSV, which is a considerable difference. The mean ratio for FN is 1.19, which also suggests a considerably large difference.
Table~\ref{table3} (b) is the summarized comparison results using cumulative log predictive Bayes factors. $\log (B_{O,PG})$ and $\log (B^{EW}_{O,PG})$ respectively denote the cumulative log predictive Bayes factor of O-DCFMSV against PG for $\by_{t+1}$ and $\bm w^\prime \by_{t+1}$. We see that $\log (B_{O,PG}) = 14.94 > 5$, suggesting very strong evidence in favor of O-DCFMSV. Similarly, for the equally-weighted portfolio, we have $\log (B^{EW}_{O,PG}) = 13.18 > 5$, which again strongly supports O-DCFMSV. In conclusion, all the evidence strongly suggests that O-DCFMSV outperforms PG's model.

\begin{figure}[ht]
\begin{center}
\includegraphics[height=2in]{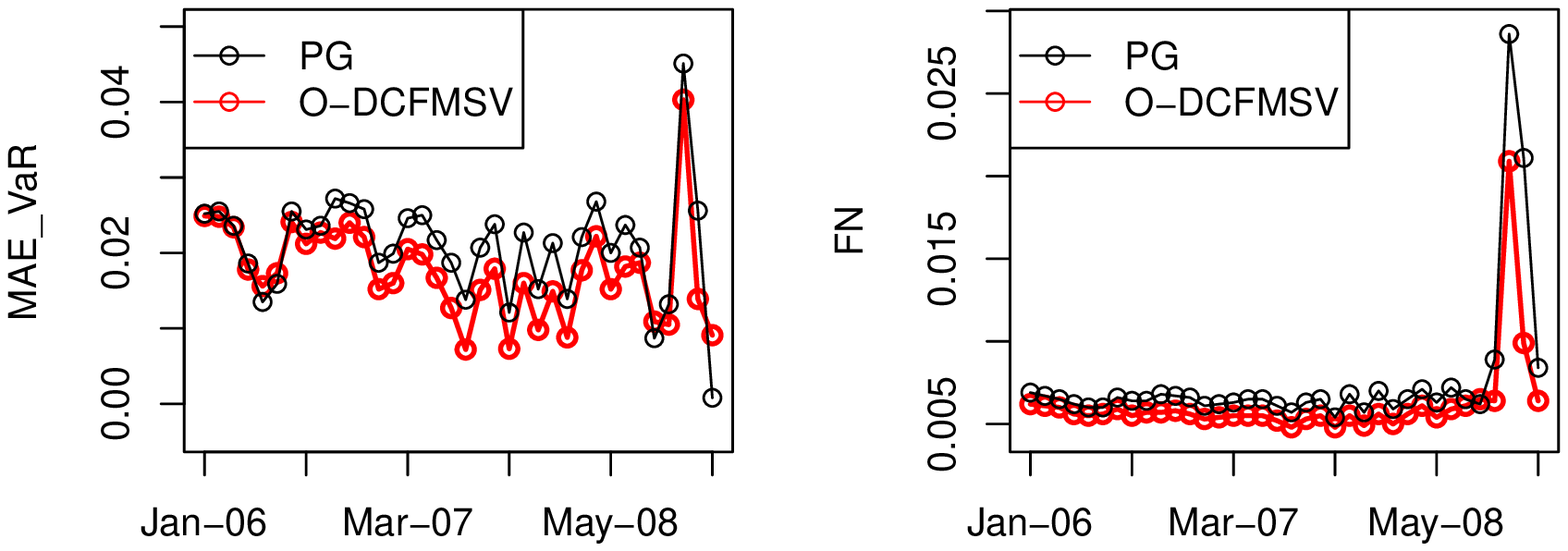}
\caption{Period-by-period comparison using $\text{MAE}_{\text{VaR}}$ and $\text{FN}$. The portfolio return data.}\label{fig4.4}\vspace{0.5in}
\includegraphics[height=2in]{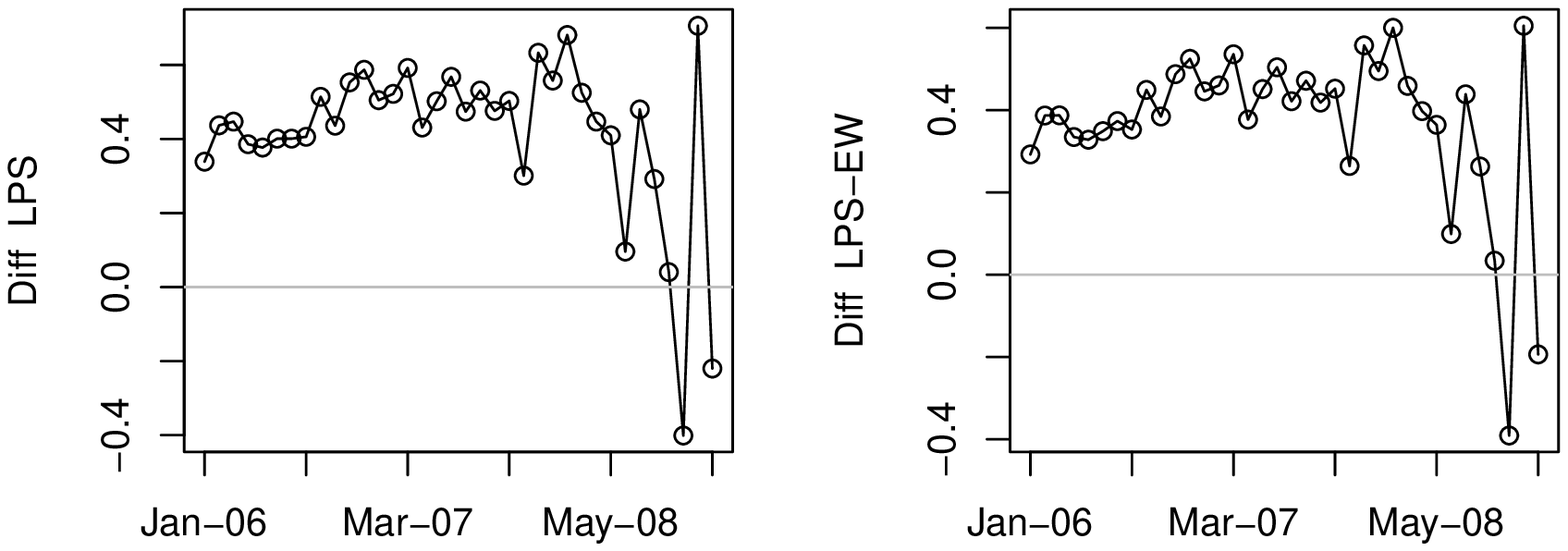}
\caption{Difference of the predictive log-likelihood for the returns and for the equally-weighted portfolio returns. The portfolio return data.}\label{fig4.5}
\end{center}
\end{figure}

The results in Table~\ref{table3} (a) and (b) are aggregated over the prediction period and do not show how the two models perform at each time point. To be more convincing, we examine the ``periodwise''performance. Figure~\ref{fig4.4} shows the period-by-period results for $\text{MAE}_{\text{VaR}}$ and $\text{FN}$. It is readily seen that the values of O-DCFMSV are constantly smaller than those of PG's. Figure~\ref{fig4.5} displays the period-by-period plot of the difference in LPS and LPS-EW (O-DCFMSV minus PG). From Figure~\ref{fig4.5} we observe that the differences in LPS and LPS-EW of O-DCFMSV minus PG are constantly greater than 0, which shows that the LPS and LPS-EW of O-DCFMSV are constantly larger than those of PG's model. The only noticeable drop occurs at October 2008, which is the month right after the bankruptcy filing of Lehman Brothers and the bailout of Fannie Mae and Freddie Mac. The market was extremely volatile at that time. In fact, according to the result not shown here, the values of the predictive density functions for both models are only about $\exp(-87)$ to $\exp(-86)$, which suggests that in such an extreme environment, both models appear to be equally unlikely (i.e., they do not hold). Consequently, the result at this time point is arguably unrepresentative.
Overall, based on these results, we conclude that O-DCFMSV generally performs better than the PG model in terms of one-step-ahead prediction.

\begin{table}
\caption{Results for the comparison of O-DCFMSV with PG's model using F-F portfolio data. $\mbox{R-MAE}_{\mbox{VaR}}$ and $\mbox{R-FN}$ are the ratios of PG to O-DCFMSV in $\mbox{MAE}_{\mbox{VaR}}$ and FN, respectively.}\label{table3}
\vspace{0.5cm}
\begin{minipage}[b]{0.5\linewidth}
\centering
(a) Empirical error-based measures.\vspace{0.2cm}
\begin{tabular}{lr}
\hline
Measure & Mean \\
\hline
$\mbox{R-MAE}_{\mbox{VaR}}$  & 1.218 \\[2pt]
$\mbox{R-FN}$                & 1.191 \\[2pt]
\hline
\end{tabular}
\end{minipage}
\begin{minipage}[b]{0.5\linewidth}
\centering
(b) cumulative log predictive Bayes factor.\vspace{0.2cm}
\begin{tabular}{lr}
\hline
Measure & Value \\
\hline
$\log (B_{O,PG})$      &  14.940 \\[2pt]
$\log (B^{EW}_{O,PG})$ &  13.176 \\[2pt]
\hline
\end{tabular}
\end{minipage}
\end{table}

\subsection{Example 2 -- Stock Return Data}\label{dailycase}
The second example fits the monthly stock return data. We collect 20 historical stock prices from Yahoo! Finance.
The observation period is January 1977 - June 2007, with 366 observations. We calculate the stock returns by taking $\log P_{t,j} - \log P_{t-1,j},$ $t=2,...,366$ and $j = 1,...,20$, where $P_{t,j}$ is the price of the $j$th stock at time $t$. This generates a set of return data with a sample size $T=365$. Similar to the example given in \cite{philipov:glickman:2006b}, in this illustration we model the stock returns using two pairs of factors, (Mkt, SMB) and (Mkt, HML), respectively. We also compare O-DCFMSV to PG based on the one-step-ahead prediction quality. The out-of-sample period is again January 2006 -- December 2008. Notice that in this case we have 20 stock returns, but during the sample period not every month has at least 20 transaction days; for this reason, unlike Example 1, here we do not calculate the empirical covariance error-based measures.

\begin{figure}[ht]
\begin{center}
\includegraphics[height=2in]{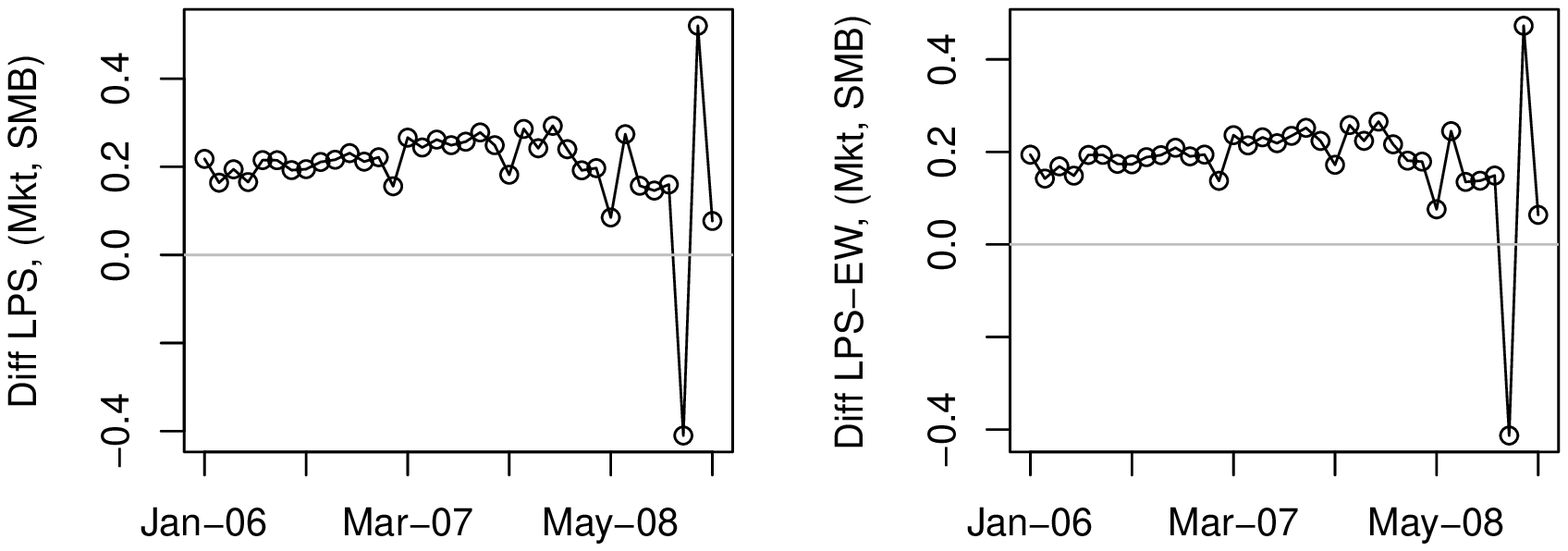}
\caption{Difference of the predictive log-likelihood for the returns and for the equally-weighted portfolio returns. (Mkt, SMB). The stock return data.}\label{fig4.6}\vspace{0.5in}
\includegraphics[height=2in]{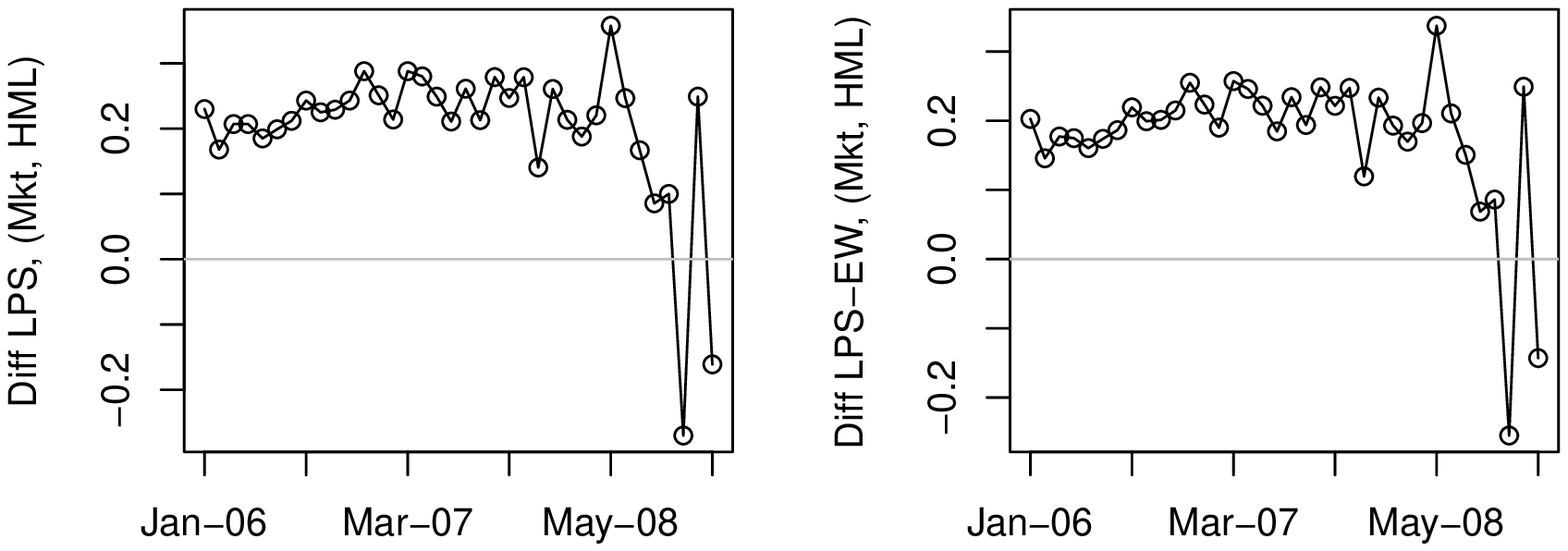}
\caption{Difference of the predictive log-likelihood for the returns and for the equally-weighted portfolio returns. (Mkt, HML). The stock return data.}\label{fig4.7}
\end{center}
\end{figure}

Table~\ref{table4} shows the aggregate results, from which we can see that, no matter what pair of factors is used, the cumulative log predictive Bayes factors $\log (B^{EW}_{O,PG})$ and $\log (B_{O,PG})$ are both greater than 5, suggesting very strong evidence in favor of the O-DCFMSV model. Figure~\ref{fig4.6} and Figure~\ref{fig4.7} respectively display the period-by-period differences in LPS and LPS-EW (O-DCFMSV minus PG) for the pairs (Mkt, SMB) and (Mkt, HML). As one can see, similar to what we observe in Example 1, the differenced values are uniformly greater than 0 in both cases except the one at October 2008. The reason is the same. At this month, no matter what pair of factors is used, the values of the predictive density functions for both models are as low as $\exp(-102)$ to $\exp(-101)$. Clearly, both models fail to work under such a market condition.
Thus, again, we argue that the differences obtained at this time point may not be meaningful. Regardless of this outlier, O-DCFMSV performs uniformly better over the out-of-sample period.

\begin{table}
\caption{Results for the comparison of O-DCFMSV with PG using stock return data.}\label{table4}
\begin{center}
\begin{tabular}{lrr}
  \hline
Measure & (Mkt,SMB) & (Mkt,HML)\\
  \hline
  $\log (B_{O,PG})$        & 7.200 & 7.272 \\[2pt]
  $\log (B^{EW}_{O,PG})$   & 6.408 & 6.480 \\[2pt]
   \hline
\end{tabular}
\end{center}
\end{table}

Another natural question to ask is which of the two combinations of factors provides a better explanation to the data. This is a model selection
question which in its generality asks how many and which factors should be used and is not discussed in
\cite{philipov:glickman:2006b}. Our solution is straightforward. We can simply compare the predictive performance of the candidate models using the cumulative log predictive Bayes factor. For instance, in this illustration we have two O-DCFMSV models, one with the factors (Mkt, SMB), denoted by MS, and the other with (Mkt, HML), MH. Table \ref{table5} summarizes the comparison result of the two models in terms of $\log (B_{MS,MH})$ and $\log (B^{EW}_{MS,MH})$. We see that, both of the cumulative log predictive Bayes scores show positive (but not strong) evidence in favor of (Mkt,SMB); therefore, we may conclude that, for a two-factor O-DCFMSV model, (Mkt,SMB) is a better choice for the data. Furthermore, if we have models that contain different numbers of factors, we can also use this approach to select the ``best'' model or the optimal number of factors.

\begin{table}
\caption{The model comparison result using stock return data. The models are O-DCFMSV given (Mkt,SMB) and (Mkt,HML) as the covariates.}\label{table5}
\begin{center}
\begin{tabular}{lr}
\hline
Measure & Value\\
\hline
$\log (B_{MS,MH})$        & 1.269 \\[2pt]
$\log (B^{EW}_{MS,MH})$   & 1.187 \\[2pt]
\hline
\end{tabular}
\end{center}
\end{table}

\section{SV on the Errors}\label{S : sv errors}
It is commonly seen in financial studies that daily and weekly data exhibit more volatility than monthly data.
For this reason, we may consider allowing each of the idiosyncratic errors to follow an independent SV process,
see, e.g. \cite{pitt:shephard:1999} and \cite{chib:nardari:shephard:2006}. We modify the model form \eqref{eq: obs eqn} as
\begin{equation}\label{SVerr}
\by_t = \bB \bmf_t + \be_t, \quad \be_t = \bm \Lambda_t^{1/2} \bm u_t,
\end{equation}
where $\bmf_t$ and $\be_t$ are independent and $\bm u_t \sim \text{N}_p(\bm 0, \bm I)$. The scaling matrix $\bm \Lambda_t = \text{diag}\left(e^{h_{t1}},...,e^{h_{tp}}\right)$, where $\{h_{tj}, j=1,...,p\}$ is the log-volatility of the error terms following the SV process:
$$
h_{tj} = \mu_j + \phi_j (h_{tj}-\mu_j) + \eta_{tj}, \quad \eta_{tj}\stackrel{iid}{\sim}\text{N}(0, \sigsq_{\eta,j}).
$$
Note that, in previous sections we use the index $i=1,...,q$ for the SV processes of the factor volatilities. Here, for the log-volatilities of the errors, we use the index $j = 1,...,p$. With the specification \eqref{SVerr} we need to change the sampling scheme for $\bB$.
Following \cite{geweke:zhou:1992} and \cite{lopes:west:2004}, we set the priors
$\bm b_j \sim \mbox{N}_q(\bm 0, c_{0j}^2 \bm I_q)$, where $\bm b_j$ is the $j$th row of $\bB$. Throughout we choose $c_{0j} = 5$, which defines an uninformative prior for each $\bm b_j$.
Let $\by_j$ be the $j$th column of $\bY$ and $\bh_j = \{h_{1j},...,h_{tj}\}$. Given the likelihood function:
$$
\mathcal{L}(\bm b_j|\by_j, \bh_j, \bF) \propto \exp\left[-\frac{1}{2}(\by_j - \bF \bm b_j)^\prime\bm \Lambda_j^{-1} (\by_j - \bF \bm b_j)\right],
$$ where $\bm \Lambda_j = \text{diag}\left(e^{h_{1j}},...,e^{h_{tj}}\right)$. The conditional posterior for $\bm b_j$ is given by:
$$
P(\bm b_j|\text{rest}) \propto \exp\bigg\{ -\frac{1}{2\sigma_j^2}\left[(\bm b_j - \bm \mu_{b_j})^\prime\bSigma_{b_j}^{-1}(\bm b_j - \bm \mu_{b_j})\right] \bigg\},
$$
where $\bSigma_{b_j} = (c_{0j}^{-2}I + \bF^\prime \bm \Lambda^{-1}_j \bF)^{-1}$ and $\bm \mu_{b_j} = \bSigma_{b_j}\bF^\prime \bm \Lambda^{-1}_j \by_j$. For $\{\mu_j, \phi_j, \sigma_{\eta,j}\}$ and $h_{tj}$, in a same manner, we use the integration sampler of \cite{kim:shephard:chib:1998} to make draws. The sampling scheme for the other parameters remains unchanged.

\begin{figure}[ht]
\begin{center}
\includegraphics[height=2.1in]{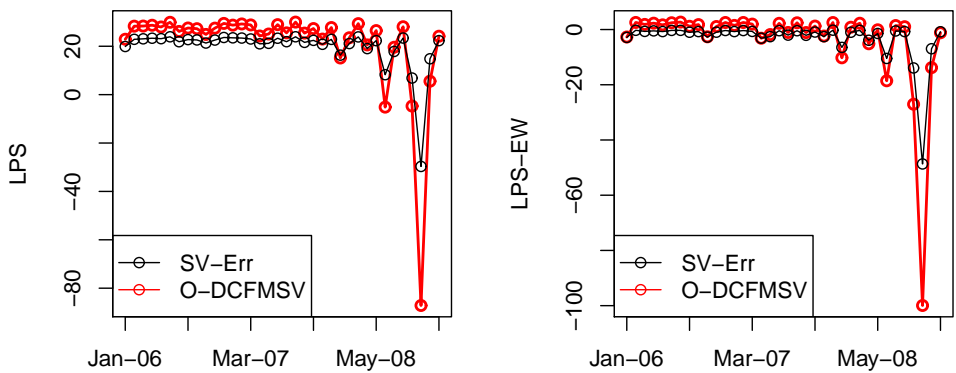}
\caption{Comparison of SV-Err and Non-SV-Err using monthly portfolio data.}\label{fig4.8}
\end{center}
\end{figure}

To know how much we gain from adding individual SV processes on the idiosyncratic error terms, using both monthly and daily data,
we compare the O-DCFMSV with SV on the errors (SV-Err) to the original model, O-DCFMSV. The model performance is compared
in terms of the cumulative log predictive Bayes factors. The first comparison is based on monthly data, where we use the same dataset as
in the first example of Section~\ref{monthlycase}. Similarly, the sample period is July 1963 -- December 2005 and the out-of-sample period is three-years long covering January 2006 -- December 2008, with a length $N=36$. Table~\ref{table6}(a) summarizes the results of the comparison.
We can see that for the monthly portfolio data, O-DCFMSV beats SV-Err in predicting $\by_{t+1}$ but not in the equally-weighted portfolio $\bm w^\prime \by_{t+1}$. However, if we examine Figure~\ref{fig4.8}, the period-by-period prediction results, we readily see that, in general O-DCFMSV performs better than or as well as SV-Err does. Nonetheless, just as we observed in Section~\ref{monthlycase}, the O-DCFMSV fails to capture the movement in returns in the single extreme period, October 2008. This failure overrides those superior performances at other periods and does not reflect the overall performance of the two models. Except for this extreme case, we should agree that O-DCFMSV suffices to model the monthly data.

In the second comparison, both models are fitted to daily data. The dataset contains 30 daily stock prices collected from Yahoo Finance. The sample period is from January 3, 2006 to July 31, 2008. We calculate the stock returns as described in Example 2. This gives us the return data of a sample size $T=649$. The out-of-sample period covers one full month, August 1 -- August 29, 2008, with a total length $N=21$. Table~\ref{table6}(b) shows the summarized result. Obviously, we see that the evidence is in favor of SV-Err since both cumulative log predictive Bayes factors are smaller than 0. The result suggests that, for daily data  the SV-Err model should be considered.

\begin{table}
\caption{Comparison results for SV-Err (S) vs. O-DCFMSV (O).}\label{table6}
\vspace{0.5cm}
\begin{minipage}[b]{0.5\linewidth}
\centering
(a) Monthly portfolio data.\vspace{0.2cm}
\begin{tabular}{lr}
\hline
Measure & Value \\
\hline
$\log (B_{O,S})$            & 41.976  \\[2pt]
$\log (B^{EW}_{O,S})$       & -27.864 \\[2pt]
\hline
\end{tabular}
\end{minipage}
\begin{minipage}[b]{0.5\linewidth}
\centering
(b) Daily stock return data.\vspace{0.2cm}
\begin{tabular}{lr}
\hline
Measure & Value \\
\hline
$\log (B_{O,S})$            & -55.512\\[2pt]
$\log (B^{EW}_{O,S})$       & -47.952\\[2pt]
\hline
\end{tabular}
\end{minipage}
\end{table}

\section{Conclusion and Discussion}\label{S: conclusion}
In this paper we propose a dynamic-correlation FMSV model where the factors are observable. The novelty is that we simultaneously allow the factors to have separate SV processes and the factor covariance process follow an inverse Wishart process, which provides great flexibility to describe the dynamics of the factors. We also develop an algorithm based on a full MCMC procedure to estimate the model. A significant advantage of the algorithm is it makes feasible to carry out prediction and model selection. This is an important improvement in this context, as it enlarges the scope of applications of the WIC-type models. From the comparisons using simulated data and various empirical data, we show that O-DCFMSV outperforms the competing model, PG's FMSV model.

Moreover, we also consider the SV-Err setting that is adopted by many authors. Our empirical result shows that, for monthly data, the basic O-DCFMSV suffices; for daily data which contain more volatility, we may need to consider the SV process on the error term. The model is easy to be extended to allow for heavy-tailedness. For example, we can add the $ad hoc$ scaled-$t$ specification suggested by \cite{kim:shephard:chib:1998} to the idiosyncratic errors. A possible future work is to allow the factors to be latent so that the model can be more flexible. When it comes to the latent factor structure, many issues need to resolve, such as the nonidentification problem, the choice of the number of factors, and so on. We expect further research on this direction.

\bibliographystyle{asa}
\bibliography{factor}

\begin{thebibliography}{20}
\newcommand{\enquote}[1]{``#1''}
\expandafter\ifx\csname natexlab\endcsname\relax\def\natexlab#1{#1}\fi

\bibitem[{Asai and McAleer(2009)}]{asai:mcaleer:2009}
Asai, M. and McAleer, M. (2009), \enquote{The structure of dynamic correlations
  in multivariate stochastic volatility models,} \textit{Journal of
  Econometric}, 150, 182--192.

\bibitem[{Barbieri et~al.(2009)Barbieri, Chang, Dubikovsky, Fox, Gladkevich,
  Gold, and Goldberg}]{barbieri:2009}
Barbieri, A., Chang, K., Dubikovsky, V., Fox, J., Gladkevich, A., Gold, C., and
  Goldberg, L. (2009), \enquote{Modeling Value at Risk with Factor,} Tech.
  rep., MSCI Barra Research.

\bibitem[{Chib et~al.(2006)Chib, Nardari, and
  Shephard}]{chib:nardari:shephard:2006}
Chib, S., Nardari, F., and Shephard, N. (2006), \enquote{Analysis of high
  dimensional multivariate stochastic volatility models,} \textit{Journal of
  Econometrics}, 134, 341--371.

\bibitem[{Doornik(2007)}]{doornik:2007}
Doornik, J. (2007), \textit{Object-Oriented Matrix Programming Using O},
  Timberlake Consultants Press, London and Oxford, 3rd ed., www.doornik.com.

\bibitem[{Geweke and Amisano(2010)}]{geweke:amisano:2010}
Geweke, J. and Amisano, G. (2010), \enquote{Comparing and evaluating Bayesian
  predictive distributions of asset returns,} \textit{International Journal of
  Forecasting}, 26, 216--230.

\bibitem[{Geweke and Zhou(1996)}]{geweke:zhou:1992}
Geweke, J. and Zhou, G. (1996), \enquote{Measuring the pricing error of the
  arbitrage pricing theory,} \textit{Review of Financial Studies}, 9, 557--587.

\bibitem[{Gilks et~al.(1995)Gilks, Best, and Tan}]{gilks:best:tan:1995}
Gilks, W., Best, N., and Tan, K. (1995), \enquote{Adaptive rejection Metropolis
  sampling within Gibbs sampling,} \textit{Applied Statistic}, 44, 455--473.

\bibitem[{Gneiting and Raftery(2007)}]{gneiting:raftery:2007}
Gneiting, T. and Raftery, A.~E. (2007), \enquote{Strictly Proper Scoring Rules,
  Prediction, and Estimation,} \textit{Journal of the American Statistical
  Association}, 102, 359--378.

\bibitem[{Harvey et~al.(1994)Harvey, Ruiz, and
  Shephard}]{harvey:ruiz:shephard:1994}
Harvey, A., Ruiz, E., and Shephard, N. (1994), \enquote{Multivariate Stochastic
  Variance Models,} \textit{Review of Economic Studies}, 61, 247--264.

\bibitem[{Jacquier et~al.(1995)Jacquier, Polson, and
  Rossi}]{jacquier:polson:rossi:1995}
Jacquier, E., Polson, N.~G., and Rossi, P.~E. (1995), \enquote{Models and prior
  distributions for multivariate stochastic volatility,} Tech. Rep. 95-18,
  CIRANO: Scientific Series, Montreal.

\bibitem[{Kass and Raftery(1995)}]{kass:raftery:1995}
Kass, R.~E. and Raftery, A.~E. (1995), \enquote{{Bayes Factors},}
  \textit{Journal of American Statistical Association}, 90, 773--795.

\bibitem[{Kim et~al.(1998)Kim, Shephard, and Chib}]{kim:shephard:chib:1998}
Kim, S., Shephard, N., and Chib, S. (1998), \enquote{Stochastic volatility:
  Likelihood inference and comparison with ARCH models,} \textit{Review of
  Economic Studies}, 65, 361--393.

\bibitem[{Ledoit et~al.(2003)Ledoit, Santa-Clara, and
  Wolf}]{Ledoit:Santa-Clara:Wolf:2003}
Ledoit, O., Santa-Clara, P., and Wolf, M. (2003), \enquote{{Flexible
  multivariate GARCH modeling with an application to international stock
  markets},} \textit{The Review of Economics and Statistics}, 85, 735--747.

\bibitem[{Liesenfeld and Richard(2006)}]{Liesenfeld}
Liesenfeld, R. and Richard, J.-F. (2006), \enquote{{Classical and Bayesian
  Analysis of Univariate and Multivariate Stochastic Volatility Models},}
  \textit{Econometric Reviews}, 25, 335--360.

\bibitem[{Lopes and Carvalho(2007)}]{LopesCal}
Lopes, H. and Carvalho, C.~M. (2007), \enquote{{Factor stochastic volatility
  with time varying loadings and Markov switching regimes},} \textit{Journal of
  Statistical Planning and Inference}, 137, 3082--3091.

\bibitem[{Lopes and West(2004)}]{lopes:west:2004}
Lopes, H. and West, M. (2004), \enquote{Bayesian model assessment in factor
  analysis,} \textit{Statistica Sinica}, 14.

\bibitem[{Philipov and Glickman(2006a)}]{philipov:glickman:2006a}
Philipov, A. and Glickman, M.~E. (2006a), \enquote{Multivariate stochastic
  volatility via Wishart processe,} \textit{Journal of Business and Economic
  Statistics}, 24, 313--328.

\bibitem[{Philipov and Glickman(2006b)}]{philipov:glickman:2006b}
--- (2006b), \enquote{Factor multivariate stochastic volatility via Wishart
  processes,} \textit{Econometric Reviews}, 25, 311--334.

\bibitem[{Pitt and Shephard(1999)}]{pitt:shephard:1999}
Pitt, M.~K. and Shephard, N. (1999), \enquote{Time-Varying Covariances: A
  Factor Stochastic Volatility Approach,} in \textit{Bayesian Statistics}, eds.
  Bernardo, J., Berger, J., Dawid, A., and Smith, A., Oxford: Oxford University
  Press, vol.~6, pp. 169--193.

\bibitem[{Rey(2000)}]{rey:2000}
Rey, D.~M. (2000), \enquote{Time-varying Stock Market Correlations and
  Correlation Breakdown,} \textit{Financial Markets and Portfolio Management},
  14, 387--412.

\end{thebibliography}
\end{document}